\documentclass{PoS}

\newcommand{\reseteqnum}{\setcounter{equation}{0}}

\newcommand{\ovl}[1]{\overline{#1}}

\newcommand{\eqn}[1]{(\ref{#1})}

\newcommand{\pslash}{p\kern-1ex /}
\newcommand{\Dslash}{{\cal D}\kern-1.5ex /}

\title{Determination of the running coupling constant $\alpha_s$ for
$N_f=2+1$ QCD with the Schr\"odinger functional scheme}

\ShortTitle{Determination of $\alpha_s$
for $N_f=2+1$ QCD with the SF scheme}

\author{\speaker{Yusuke Taniguchi}%
\ for PACS-CS Collaboration
\\
Institute of Physics, University of Tsukuba,
Tsukuba, Ibaraki, 305-8571, Japan \\
         E-mail: \email{tanigchi@het.ph.tsukuba.ac.jp}}

\abstract{
  We present an evaluation of the running coupling constant and the
  quark mass renormalization factor for $N_f=2+1$ QCD.
 The Schr\"odinger functional scheme is used as the intermediate
 scheme to carry out non-perturbative running from the low energy region,
 where physical input is introduced, to deep in the high energy perturbative
 region, where conversion to the ${\ovl{\rm MS}}$ scheme is safely performed. 

  For numerical simulations we adopted  Iwasaki gauge action and
 non-perturbatively improved Wilson fermion action with the clover term.
 Seven renormalization scales are used to cover from low to high energy
 region and three lattice spacings to take the continuum limit at each
 scale.

 Physical inputs are introduced from the previous $N_f=2+1$ simulation of
 the CP-PACS/JL-QCD collaboration, which covered 
 the up-down quark mass range heavier than $m_\pi\sim 500$~MeV,
 and that of PACS-CS collaboration for much lighter quark masses down 
 to $m_\pi=155$~MeV.
}

\FullConference{The XXVII International Symposium on Lattice Field Theory\\
		 July 26-31, 2009\\
		 Peking University, Beijing, China}

\begin{document}

\reseteqnum
\section{Introduction}

The strong coupling constant and quark masses constitute the 
fundamental parameters of the Standard Model.
It is an important task of lattice QCD to determine these
parameters using inputs at low energy scales such as 
hadron masses, meson decay constants and quark potential quantities. 
The results can be compared with independent determinations from high
energy experiments.

In the course of evaluating these fundamental parameters we need the
process of renormalization in some scheme.
The ${\ovl{\rm MS}}$ scheme is one of the most popular schemes, and 
hence one would like to evaluate these parameters through input of low
energy quantities on the lattice and convert it to the ${\ovl{\rm MS}}$
scheme.
A difficulty in this process is that the conversion is given only in a
perturbative expansion, and should be performed at high energy scales much 
larger than the QCD scale.
At the same time the renormalization scale $\mu$ should be kept much less
than the lattice spacing to reduce lattice artifacts, namely we require 
$\Lambda_{\rm QCD}\ll\mu\ll{1}/{a}$.
A practical difficulty of satisfying these inequalities in numerical 
simulations is called the window problem. 

The Schr\"odinger functional (SF) scheme
\cite{Luscher:1992an,Luscher:1993gh,Capitani:1998mq,DellaMorte:2004bc,Della Morte:2005kg}
is designed to resolve the window problem.
It has an advantage that systematic errors can be unambiguously
controlled.
A unique renormalization scale is introduced through the box size $L$.
A wide range of renormalization scales can be covered by the step
scaling function (SSF) technique.
The SF scheme has been applied for evaluation of the QCD coupling 
and the quark mass renormalization factor for
$N_f=0$ \cite{Luscher:1993gh,Capitani:1998mq} and $N_f=2$
\cite{DellaMorte:2004bc,Della Morte:2005kg}.

At low energy scales of $\mu\sim500$ MeV, where physical input is given,
we expect the strange quark contribution to be important in addition to
those of the up and down quarks.
Thus the aim of this proceeding is to go one step further
and evaluate the strong coupling constant and the quark mass
renormalization factor in $N_f=2+1$ QCD.
For setting the physical scale we employ two recent large-scale
$N_f=2+1$ lattice QCD simulations employing non-perturbatively O(a)
improved Wilson quark action \cite{Ishikawa:2007nn,Aoki:2008sm}.

Our goal is to evaluate the running coupling constant $\alpha_s(M_Z)$
and the renormalization group invariant (RGI) quark mass $M$.
The evaluation of $\alpha_s(M_Z)$ has been performed in
Ref.~\cite{Aoki:2009tf}, where one finds detailed explanation.
The objective for the quark mass is to derive a renormalization factor
$Z_M(g_0)$, which converts the bare PCAC mass at bare coupling $g_0$ to
the RGI mass.
The derivation proceed in the same steps as
Ref.~\cite{Capitani:1998mq,Della Morte:2005kg}, which we omit in this
proceeding.

\reseteqnum
\section{Step scaling function}

We adopt the renormalization group improved gauge
action of Iwasaki
and the improved Wilson fermion action with clover term,
whose improvement coefficient $c_{\rm SW}$ is given non-perturbatively
\cite{Aoki:2009tf}.
The twisted periodic boundary condition is set for the fermion field in
the three spatial directions with $\theta={\pi}/{5}$ for the coupling
constant and $\theta=0.5$ for the pseudo scalar density.
We adopted two different SF boundary conditions when evaluating the
coupling constant \cite{Luscher:1993gh,DellaMorte:2004bc}
and for the pseudo scalar density renormalization factor
\cite{Capitani:1998mq,Della Morte:2005kg}.

We adopt seven renormalized coupling values to cover weak
to strong coupling regions.
For each coupling we use three boxes $L/a=4, 6, 8$ to take the
continuum limit.
For three lattice sizes the values of $\beta$ and $\kappa$ are tuned to
reproduce the same renormalized coupling keeping the PCAC mass to zero.
On these parameters we evaluate the coupling constant $\ovl{g}^2(L)$,
$\ovl{g}^2(2L)$ and the renormalization factor ${Z_P(g_0,L/a)}$,
${Z_P(g_0,2L/a)}$.
Tiny deviations in the renormalized coupling $\ovl{g}^2(L)$
and the PCAC mass are corrected perturbatively \cite{Bode:1999sm}
and we have the SSF's on the lattice.

\begin{figure}
 \begin{center}
  \includegraphics[width=4.cm]{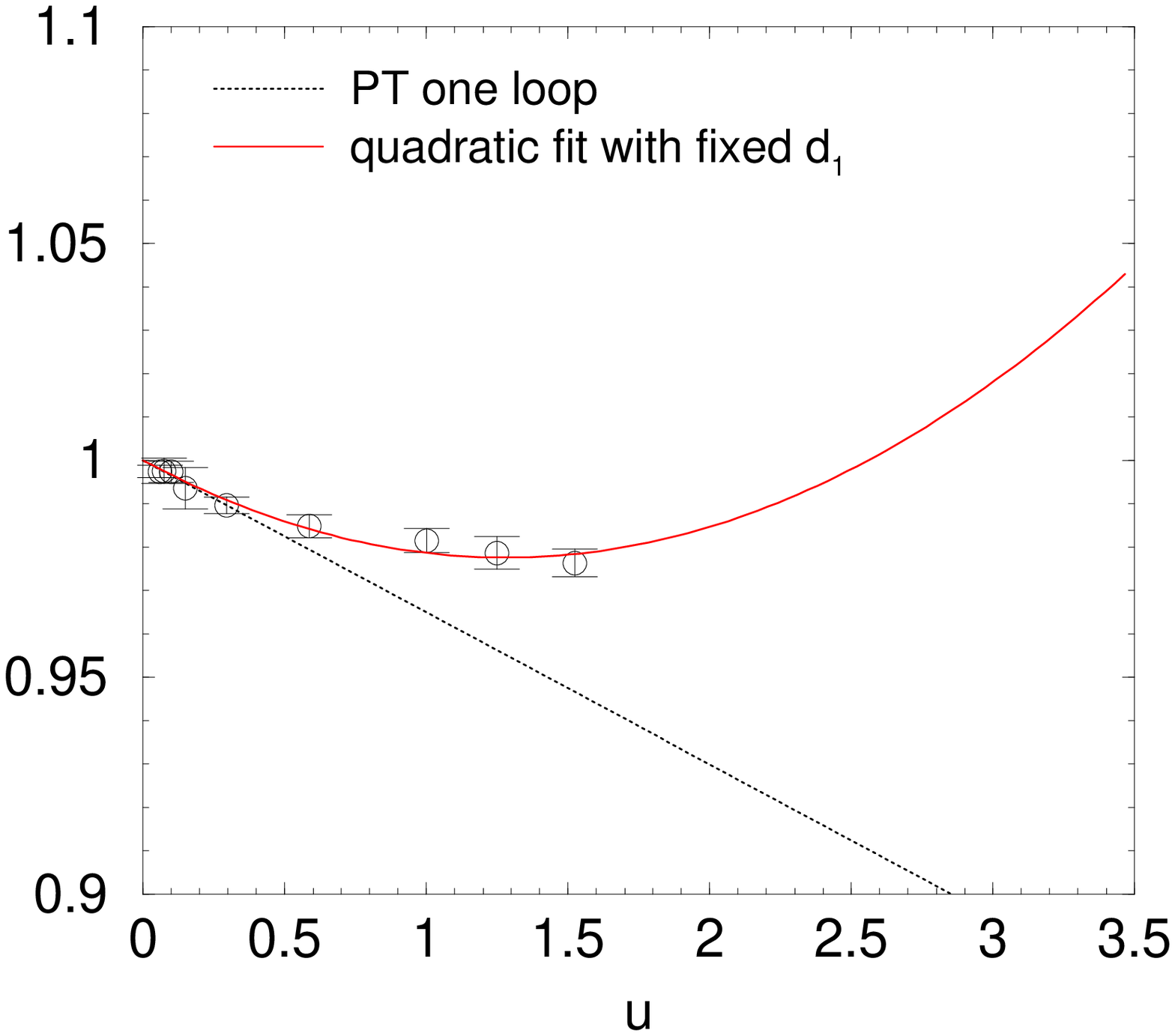}
  \includegraphics[width=4.cm]{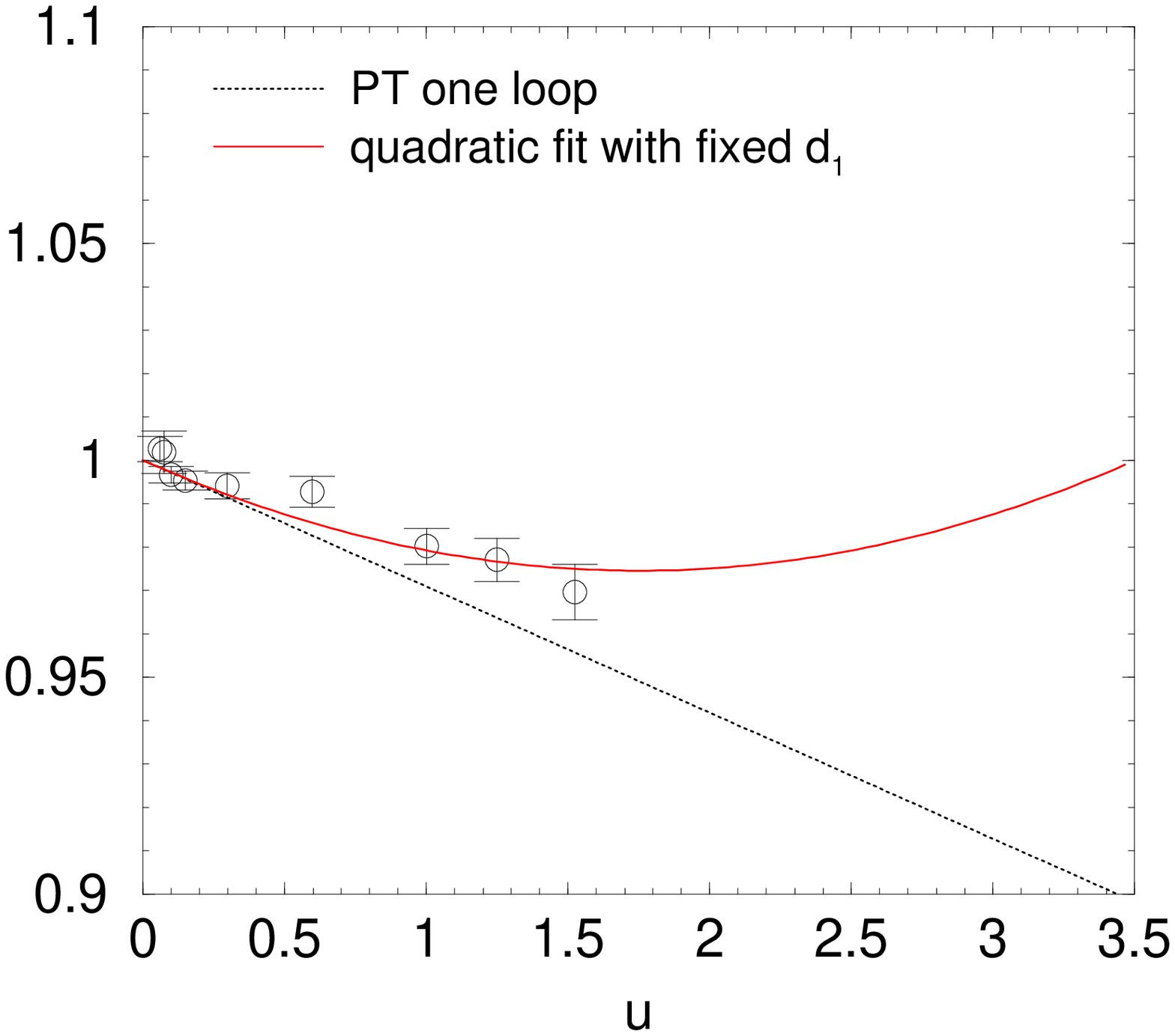}
  \includegraphics[width=4.cm]{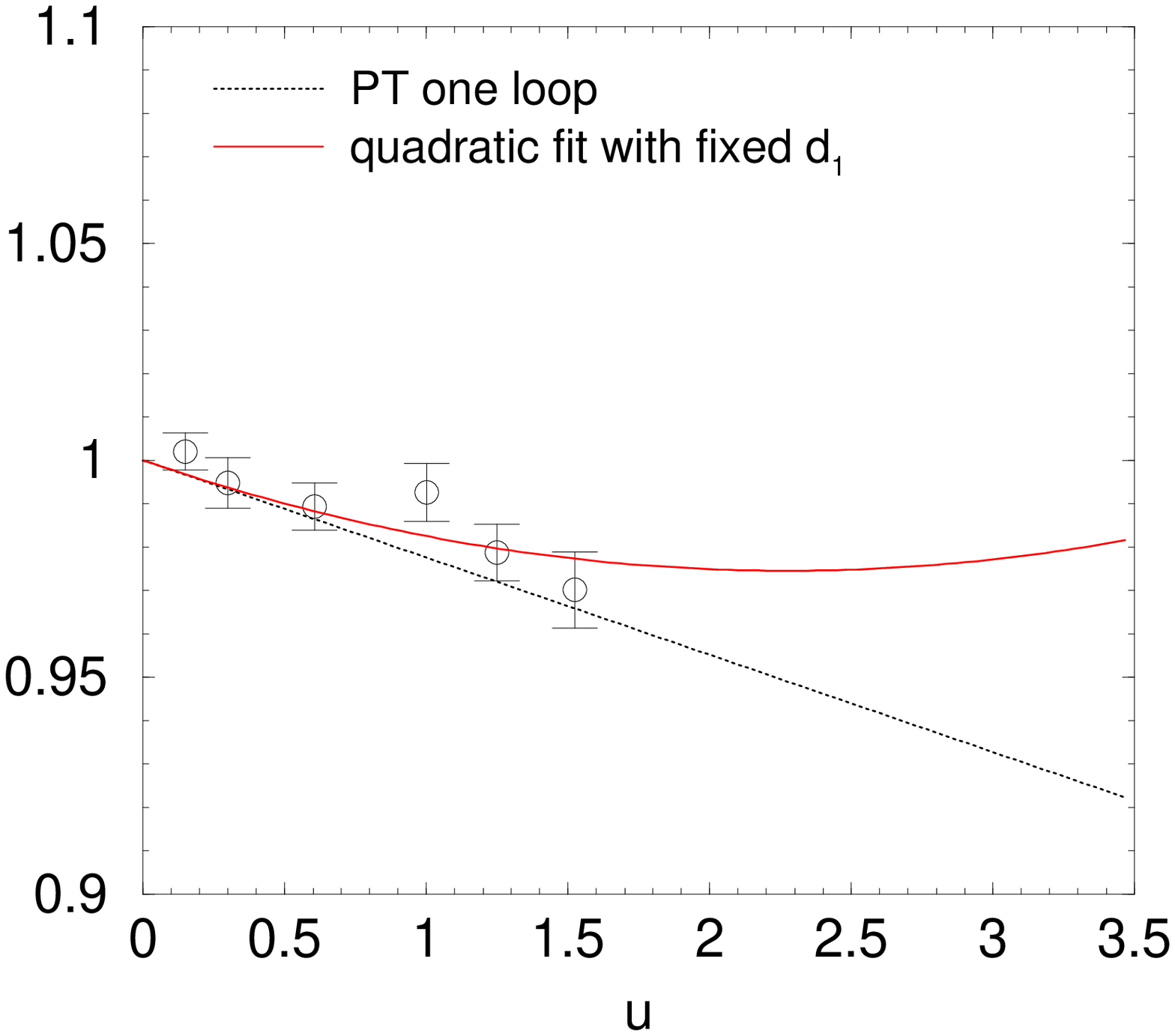}
  \includegraphics[width=4.cm]{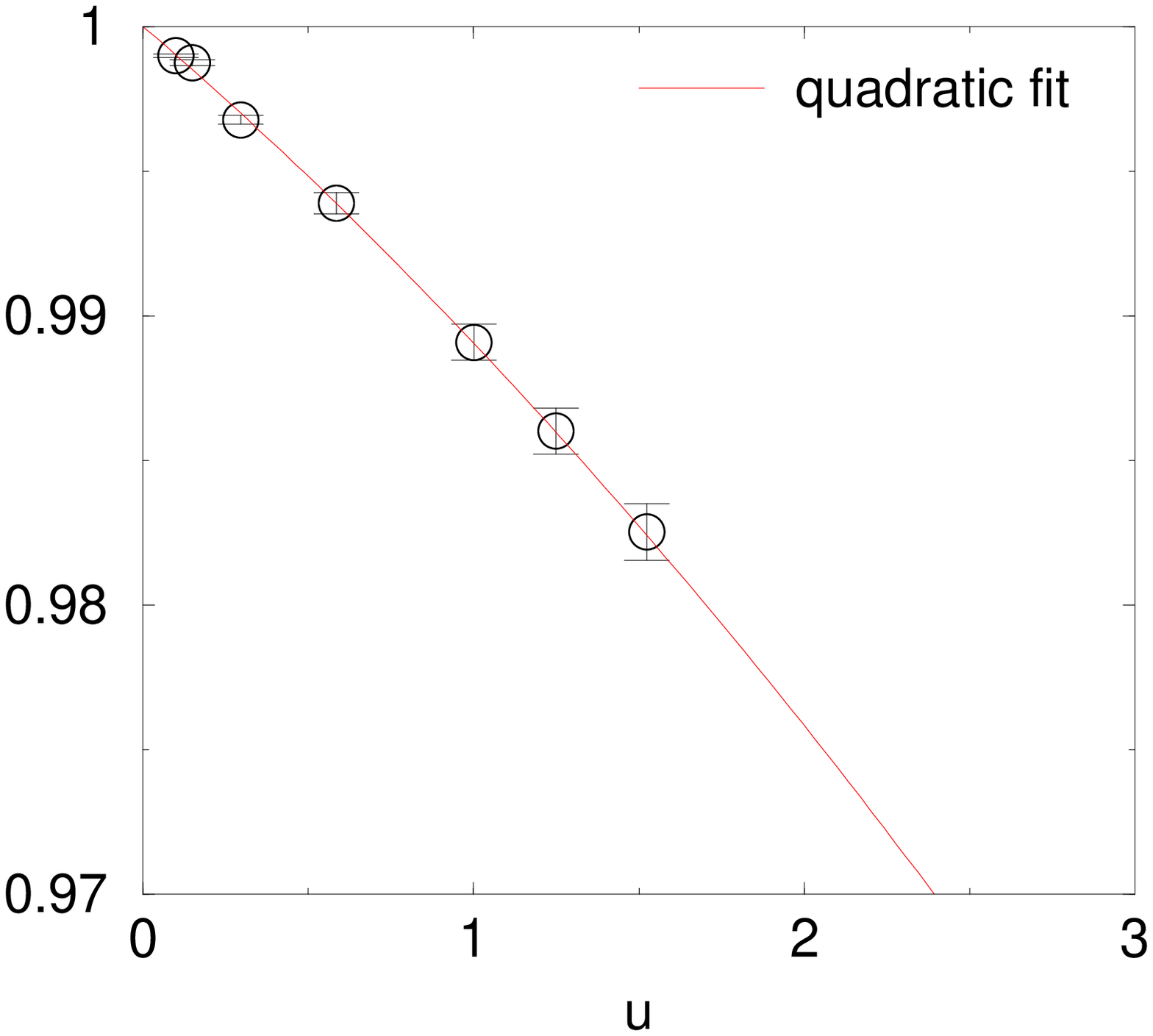}
  \includegraphics[width=4.cm]{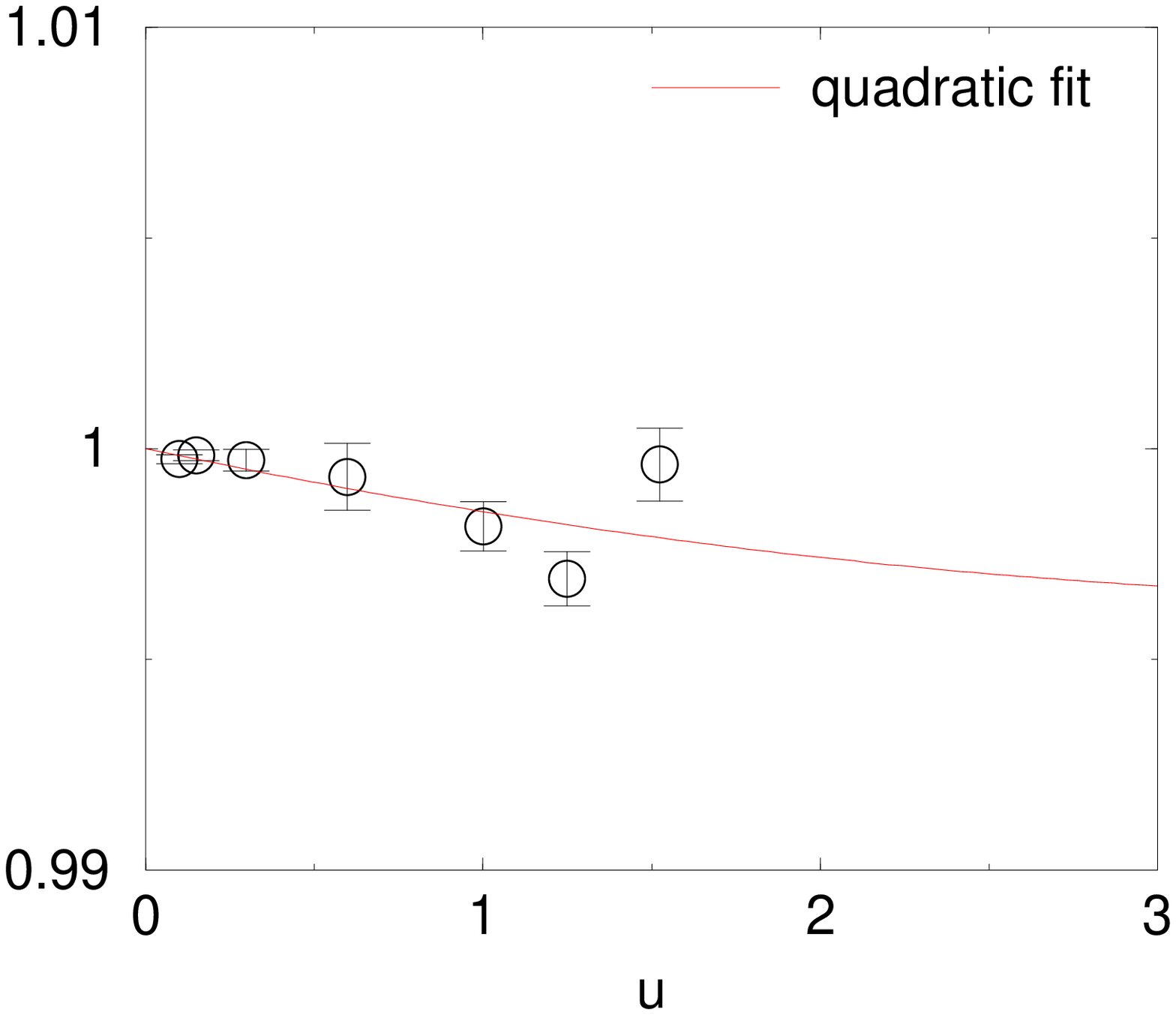}
  \includegraphics[width=4.cm]{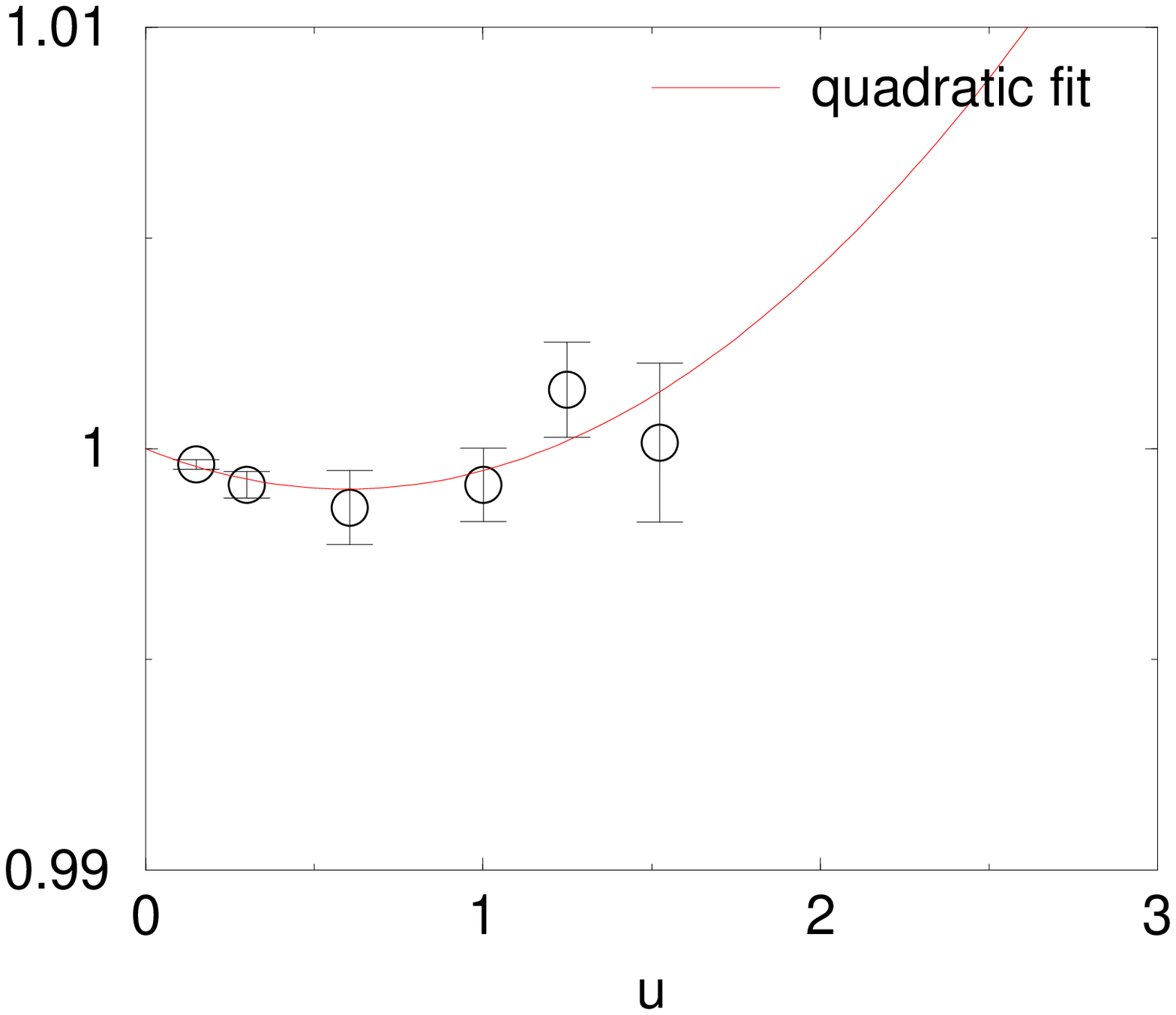}
  \caption{Polynomial fit of discrepancy
  $\Sigma\left(u,{a}/{L}\right)/\sigma^{(3)}(u)$ (upper three figures) 
  and $\Sigma_P\left(u,{a}/{L}\right)/\sigma_P^{(3)}(u)$ (lower) at high
  $\beta\gtrsim4$.
  The fit is given for each lattice spacings $a/L=1/4$ (left), $a/L=1/6$
  (middle) and $a/L=1/8$ (right).
  Black dotted line is a perturbative one loop behavior and red solid
  line is a quadratic fit.}
  \label{fig:ordera}
 \end{center}
\end{figure}

We perform a perturbative improvement of the SSF before taking the
continuum limit, for which we need an evaluation of the lattice artifact
$\delta_{(P)}(u,a/L)=
{(\Sigma_{(P)}\left(u,{a}/{L}\right)-\sigma_{(P)}(u))}/{\sigma_{(P)}(u)}$.
Although $\delta(u,a/L)$ is evaluated at one loop level its value is
rather large and is shown to be applicable only for very weak coupling
region \cite{Aoki:2009tf}, which reveals importance of two loop
coefficient.
On the other hand $\delta_P$ is not known for our setup.
Instead of calculating $\delta_P$ and $\delta$ at one/two-loop
level perturbatively we calculate SSF's directly by Monte-Carlo sampling
at very weak coupling $\beta\ge10$.
We define $\delta_{(P)}(u,a/L)$ by a deviation from the perturbative
SSF's $\sigma_{(P)}^{(3)}$ at three (two) loops order
\cite{Bode:1999sm}.
The deviation is fitted in a polynomial form for each $a/L$,
\begin{eqnarray}
1+\delta_{(P)}(u,a/L)=1+d_1^{(P)}(a/L)u+d_2^{(P)}(a/L)u^2.
\end{eqnarray}
We tried a quadratic fit using data at $u\le1.524$, which is plotted in
Fig.~\ref{fig:ordera}.
The one loop coefficient $d_1(a/L)$ is fixed to its perturbative value
for the coupling SSF.

Since the quadratic fit provides a reasonable description of data
we opt to cancel the $O(a)$ contribution dividing out the SSF by the
quadratic fit.
On the other hand the deviation is consistent with zero within one standard
deviation for $\delta_P(u,1/8)$ at $u\gtrsim1$ we do not apply an
improvement for this case.

\begin{figure}
 \begin{center}
  \includegraphics[width=5.cm]{fig/sigma.continuum3.eps}
  \includegraphics[width=5.5cm]{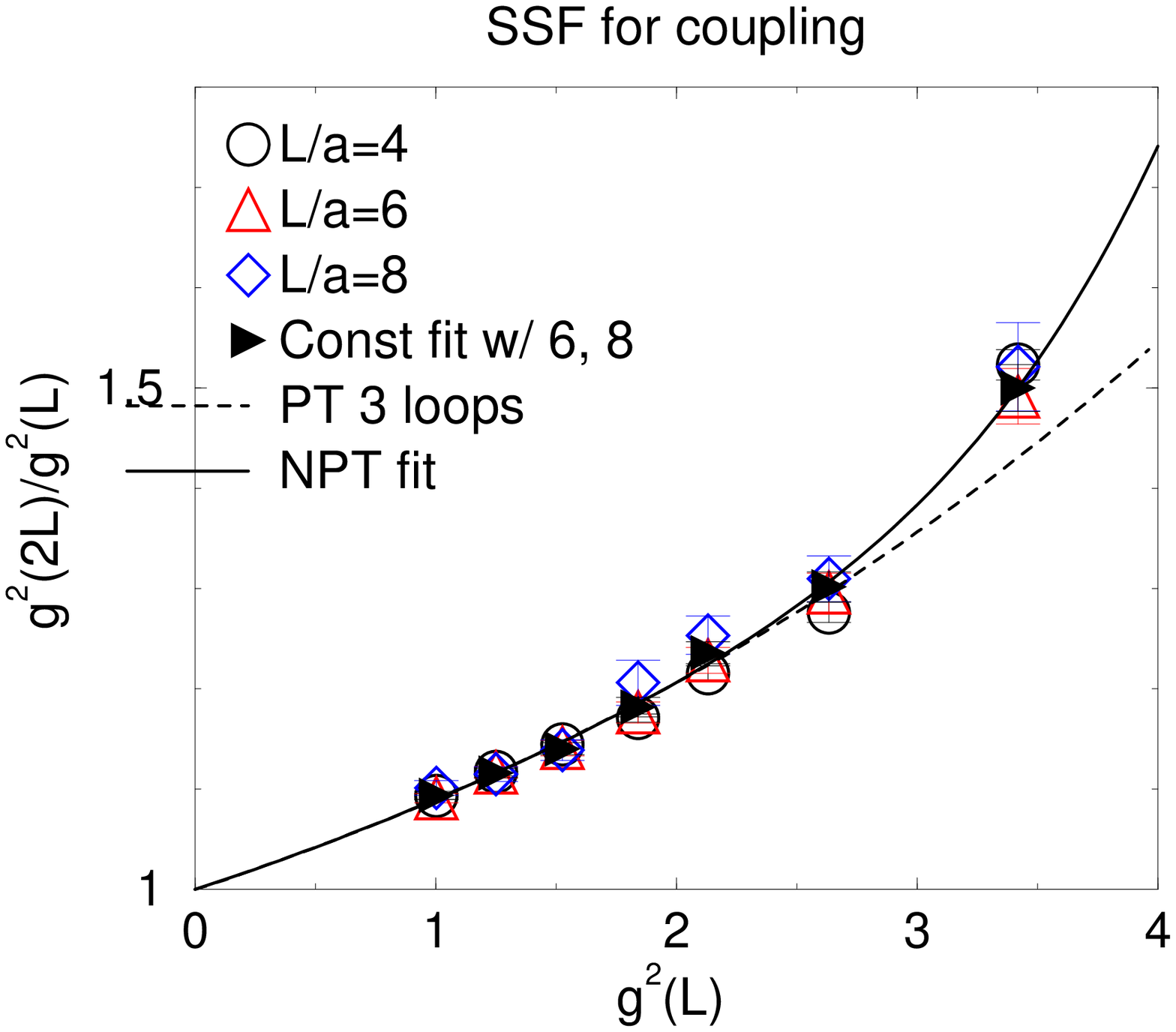}
  \caption{The SSF of the coupling constant with its continuum
  extrapolation at each renormalization scale (left).
  Right panel is a RG flow of the SSF.
  Dotted line is three loops perturbative running.
  Solid line is a polynomial fit.
  }
  \label{fig:SSF}
 \end{center}
\end{figure}

\begin{figure}
 \begin{center}
  \includegraphics[width=5.5cm]{fig/sigmap.continuum3.eps}
  \includegraphics[width=5.5cm]{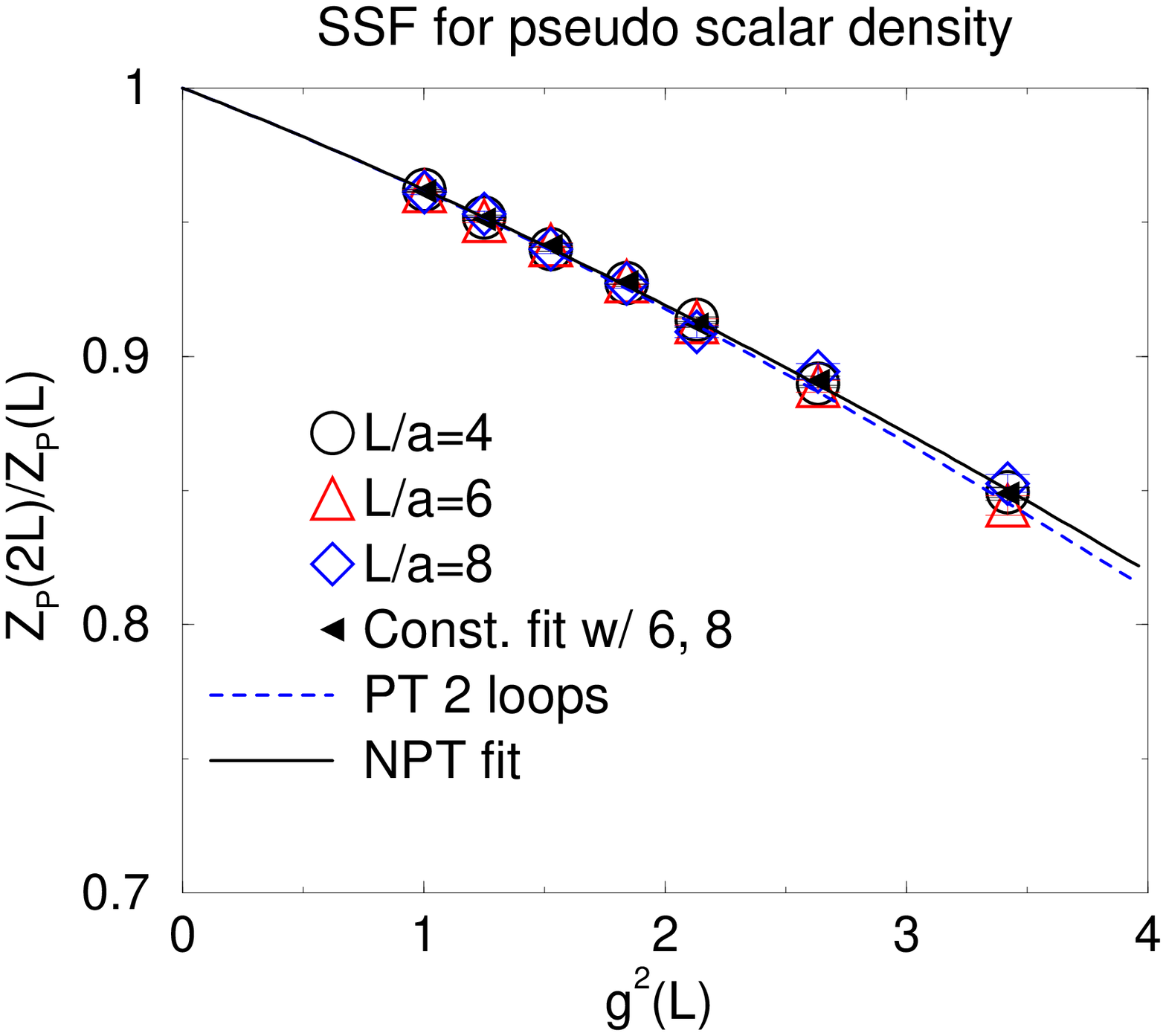}
  \caption{The SSF for the pseudo scalar density (left).
  RG flow of the SSF (right).}
  \label{fig:SSFP}
 \end{center}
\end{figure}

Scaling behavior of the improved SSF is plotted in Fig.~\ref{fig:SSF}
for the coupling constant and in Fig.~\ref{fig:SSFP} for the pseudo
scalar density.
Almost no scaling violation is found.
We performed three types of continuum extrapolation:
a constant extrapolation with the finest two (filled symbols) or all
three data points (open symbols), or a linear extrapolation with all
three data points (open circles), which are consistent with each other.
We employed the constant fit with these two data point to find our
continuum value.
The RG running of the continuum SSF is plotted in the same figure at
right panel.
The fitting functions
\begin{eqnarray}
\sigma(u)&=&u+s_0u^2+s_1u^3+s_2u^4+s_3u^5+s_4u^6,
\\
s_2&=&0.002265,\quad
s_3=-0.00158,\quad
s_4=0.000516,
\\
\sigma_P(u)&=&1+p_0u+p_1u^2+p_2u^3,
\\
p_1&=&-0.002861\quad
p_2=0.000093.
\end{eqnarray}
are also plotted (solid line) together with the three/two loops
perturbative running (dashed line), where $s_0$, $s_1$ and $p_0$ are set
to their perturbative values.

\reseteqnum
\section{Introduction of physical scale}

CP-PACS and JLQCD Collaborations jointly performed an $N_f=2+1$
simulation with the $O(a)$ improved Wilson action and the Iwasaki gauge
action \cite{Ishikawa:2007nn}.
Three values of $\beta$, $1.83$, $1.90$ and $2.05$ were adopted to take
the continuum limit and the up-down quark mass covered a rather heavy
region corresponding to $m_\pi/m_\rho=0.63-0.78$.
This project has been taken over by the PACS-CS Collaboration aiming at
simulations at the physical light quark masses
\cite{Aoki:2008sm}, where
results at a single lattice spacing $\beta=1.90$ is available 
with very light quark masses down to $m_\pi/m_\rho\approx 0.2$.

We adopt those results to introduce the physical scale into the present
work through the low energy reference scale $L_{\rm max}$ in MeV units.
We employ the hadron masses $m_\pi$, $m_K$, $m_\Omega$ as inputs 
and use the lattice spacing $a$ as an intermediate scale.

We evaluate the renormalized coupling and the pseudo scalar density
renormalization factor at the same $\beta$ in the chiral limit.
The reference scale $L_{\rm max}$ is given by the box size we adopt in
this evaluation.
The renormalized coupling $\ovl{g}^2(L_{\rm max})$ should
not exceed our maximal value $5.13$ of the SSF very much.
The values of the coupling constant and the renormalization factor
$Z_P(L_{\rm max})$
are listed in Table \ref{tab:lmax}.
\begin{table}[htb]
\begin{center}
\begin{tabular}{|c|c|c|c|c|}
\hline
$\beta$ & $\kappa$ & $L_{\rm max}/a$ & $\ovl{g}^2(L_{\rm max})$
 & $Z_P(L_{\rm max})$ \\
\hline
$1.83$ & $0.13608455$ & $4$ & $5.565(54)$ & $0.57519(32)$ \\
\hline
$1.90$ & $0.1355968$ & $4$ & $4.695(23)$ & $0.60784(27)$ \\
\hline
$2.05$ & $0.1359925$ & $6$ & $4.740(79)$ & $0.56641(44)$ \\
\hline
\end{tabular}
 \caption{The renormalized coupling and the renormalization factor $Z_P$
 at $\beta=1.83$, $1.90$, $2.05$ to define the reference scale
 $L_{\rm max}$.}
\label{tab:lmax}
\end{center}
\end{table}

We calculate the axial vector current renormalization factor according
to the procedure in Ref.~\cite{Luscher:1996jn,Della Morte:2005rd}.
We adopt the renormalization condition \cite{Della Morte:2005rd},
which is applicable to non-vanishing PCAC mass, with connected diagrams
only.
The physical box size is fixed to approximately same value $L\sim0.75$
fm.
Since we did not find any significant $\theta$ dependence we evaluate
the renormalization factors at $\theta=0.5$.
Preliminary results are listed in table \ref{tab:za}.
\begin{table}[htb]
\begin{center}
\begin{tabular}{|c|c|c|c|c|c|c|c|c|c|c|c|}
\hline
$\beta$ & $\kappa$ & size & $m_{\rm AWT}$ & $Z_V$ & $Z_A$ \\
\hline
$1.83$ & $0.138466$ & $6^3\times18$ & $-0.0003(19)$
 & $0.751(26)$ & $0.965(23)$\\
\hline
$1.90$ & $0.137556$ & $8^3\times18$ & $0.00150(63)$
 & $0.7424(63)$ & $0.8596(77)$\\
\hline
$2.05$ & $0.136116$ & $12^3\times30$ & $0.00291(38)$
 & $0.7717(41)$ & $0.8117(54)$\\
\hline
\end{tabular}
 \caption{The (axial) vector current renormalization factor at
 $\beta=1.83$, $1.90$, $2.05$.}
\label{tab:za}
\end{center}
\end{table}

\reseteqnum
\section{Strong coupling constant at $M_Z$ and RGI mass renormalization
 factor}

We derive the strong coupling constant $\alpha_s(M_Z)$ and
$\Lambda_{\ovl{\rm MS}}^{(5)}$ according to the procedure in
Ref.~\cite{Aoki:2009tf}.
The results are listed in Table \ref{tab:alphas}.
\begin{table}[htb]
\begin{center}
\begin{tabular}{|c|c|c|c|c|c|}
\hline
$\beta$ & $\alpha_s(M_Z)$ & $\Lambda_{\ovl{\rm MS}}^{(5)}$ (MeV)
 & $Z_M$ & $Z_m^{\ovl{\rm MS}}(2\ {\rm GeV})$ \cr
\hline
$1.83$ & $0.1208(13)$ & $243(17)$ & $2.084(55)$ & $1.616(41)$ \cr
$1.90$ & $0.1206(14)$ & $240(18)$ & $1.870(25)$ & $1.446(17)$ \cr
$2.05$ & $0.1198(16)$ & $231(20)$ & $1.888(26)$ & $1.446(15)$ \cr
\hline
$1.90$ & $0.1225(14)$ & $266(20)$ & $1.870(25)$ & $1.484(18)$ \cr
\hline
\end{tabular}
\caption{The strong coupling $\alpha_s(M_Z)$ and the RGI
 scale $\Lambda_{\ovl{\rm MS}}^{(5)}$ for five flavors.
 Also listed are preliminary results of $Z_M$ for the RGI mass and
 $Z_m^{\ovl{\rm MS}}$ in the $\ovl{\rm MS}$ scheme.
 The last row for $\beta=1.90$ is given by an input from
 Ref.~\cite{Aoki:2008sm}.
}
\label{tab:alphas}
\end{center}
\end{table}
The error includes the statistical error of the renormalized couplings
in addition to the statistical error of the
lattice spacing.
The experimental errors of $m_c$, $m_b$ and $M_Z$ are also included.
Preliminary results of the renormalization factor $Z_M$ for the RGI
mass are also given in Table \ref{tab:alphas} together
with the mass renormalization factor
$Z_m^{\ovl{\rm MS}}(\beta,\mu=2\ {\rm GeV})$ in the $\ovl{\rm MS}$
scheme.

\begin{figure}[htbp]
\begin{center}
 \includegraphics[width=5.cm]{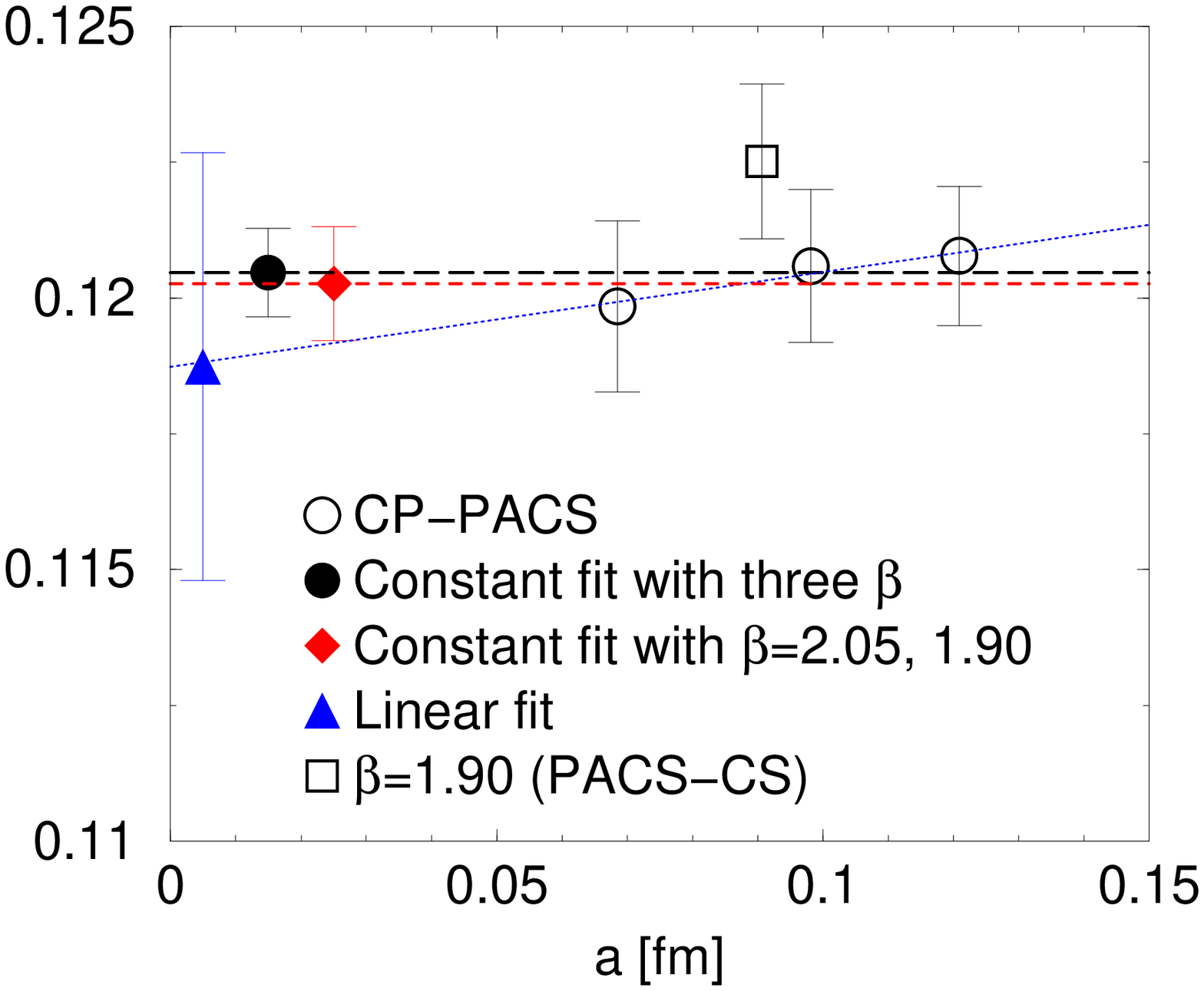}
 \includegraphics[width=5.cm]{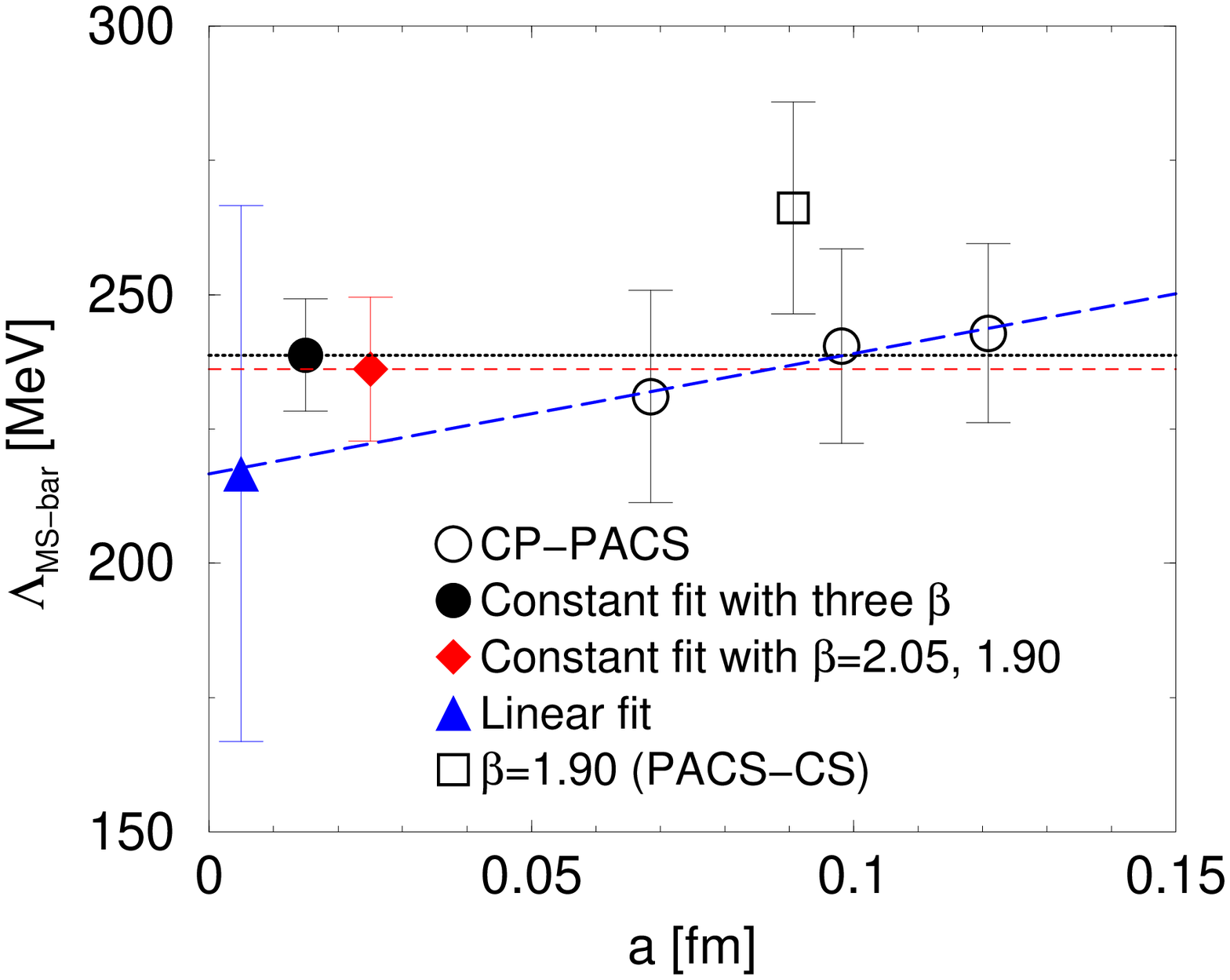}
\\
 \includegraphics[width=5.cm]{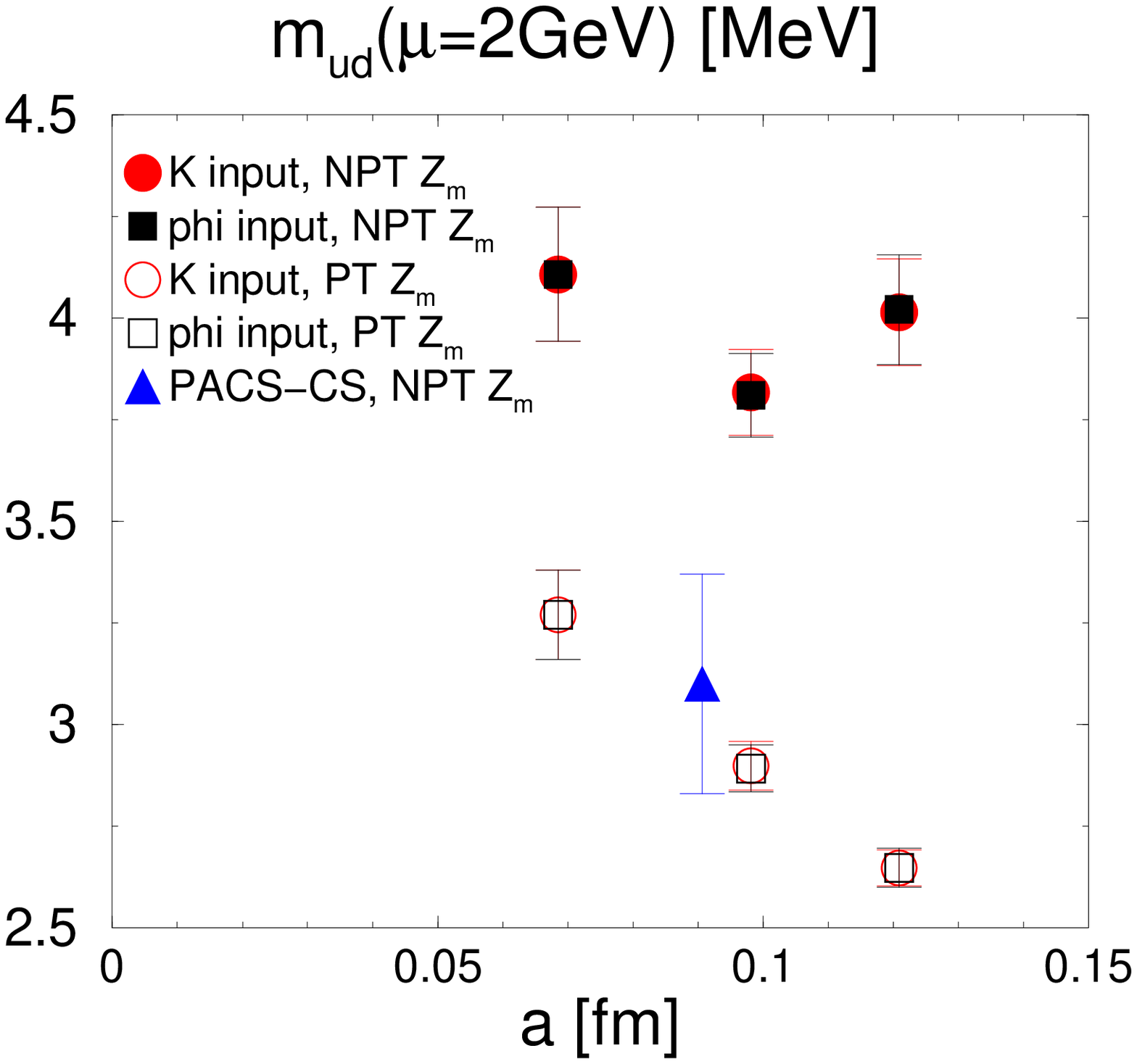}
 \includegraphics[width=5.cm]{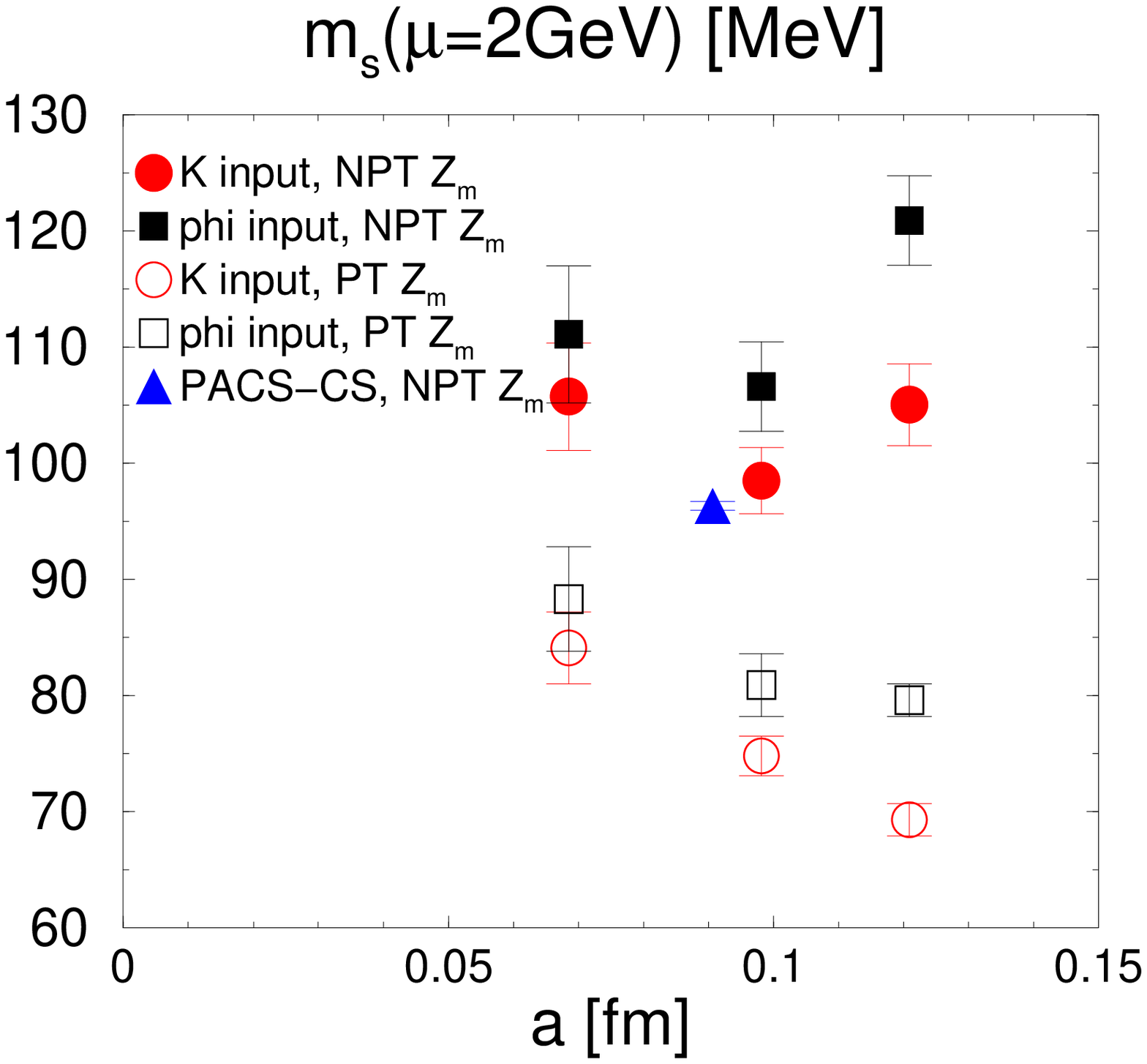}
 \caption{Scaling behavior of $\alpha_{\ovl{\rm MS}}(M_Z)$ (upper left)
 and $\Lambda_{\ovl{\rm MS}}^{(5)}$ (upper right).
 Scaling behavior of $m_{ud}^{\ovl{\rm MS}}$ (lower
 left) and $m_{s}^{\ovl{\rm MS}}$ (lower right); filled circles and
 squares are from \cite{Ishikawa:2007nn}, filled triangle is from
 \cite{Kuramashi}, open symbols are perturbatively renormalized masses
 \cite{Ishikawa:2007nn}.}
\label{fig:alphamz}
\end{center}
\end{figure}

As the last step we take the continuum limit using the three lattice
spacings from Ref.~\cite{Ishikawa:2007nn}.
The scaling behavior of $\alpha_s(M_Z)$ and
$\Lambda_{\ovl{\rm MS}}^{(5)}$ is plotted in Fig.~\ref{fig:alphamz}
together with that from latest input \cite{Aoki:2008sm}.
Two results agree with each other at the same $\beta=1.90$.
We tested three types of continuum extrapolation, which agree with each
other and we adopt the constant fit with three data points for our final
results:
\begin{eqnarray}
\alpha_s(M_Z)=0.12047(81)(48)(^{+0}_{-173}),\quad
\Lambda_{\ovl{\rm MS}}^{(5)}=239(10)(6)(^{+0}_{-22}) \;{\rm MeV},
\label{eqn:alpha5}
\end{eqnarray}
where the first parenthesis is statistical error and the second is
systematic error of perturbative matching of different flavors.
The last parenthesis is a difference between the constant and a linear
extrapolation and is a systematic error due to finite lattice spacing for
physical inputs.

The results from the physical inputs of our latest
Ref.~\cite{Aoki:2008sm} are given by 
\begin{eqnarray}
\alpha_s(M_Z)=0.1225(14)(5),\quad
\Lambda_{\ovl{\rm MS}}^{(5)}=266(20)(7) \;{\rm MeV}.
\end{eqnarray}
Difference between the two physical inputs may reflect mainly a
systematic error due to chiral extrapolation toward light quark masses,
with the assumption the scaling violation is small also in the latter
case.

We also plot preliminary scaling behavior of the light quark masses
renormalized at $\mu=2$ GeV in $\ovl{\rm MS}$ scheme together with
perturbatively renormalized masses \cite{Ishikawa:2007nn}.

\reseteqnum
\section{Conclusion}

We have presented a calculation of the running coupling constant and the
quark mass renormalization factor for the $N_f=2+1$ QCD in the mass
independent Schr\"odinger functional scheme in the chiral limit.
With the ``perturbative'' improvement the SSF's shows good scaling
behavior and the continuum limit seems to be taken safely
with a constant extrapolation of the finest two lattice spacings.

With the non-perturbative renormalization group flow we are able to
estimate $\alpha_s(M_Z)$ and the quark mass renormalization factor with
some physical inputs for energy scale.
The physical scale is introduced from the recent spectrum simulations
\cite{Ishikawa:2007nn,Aoki:2008sm} through the hadron
masses $m_\pi$, $m_K$, $m_\Omega$.
Our coupling constant \eqn{eqn:alpha5} in the continuum limit is
consistent with recent lattice results and the Particle Data Group
average $\alpha_s(M_Z)=0.1176(20)$.


\end{document}